\documentclass[12pt]{article}

\usepackage{amsmath}
\usepackage{amssymb}
\usepackage{graphicx}
\usepackage{dsfont}

\begin{document}

\title{Holonomy reduced dynamics of triatomic molecular systems}

\author{\"{U}nver \c{C}ift\c{c}i$^1$\thanks{uciftci@nku.edu.tr} and Holger Waalkens$^2$\thanks{h.waalkens@rug.nl}}

\maketitle

\noindent
{\small $^1$ Department of Mathematics, Nam\i k Kemal University, 59030, Tekirda\u{g}, Turkey}\\
\noindent
{\small $^2$ Johann Bernoulli Institute for Mathematics and Computer Science, University of Groningen, PO Box 407, 
9700 AK Groningen, The Netherlands}



\begin{abstract}
Whereas it is easy to reduce the translational symmetry of a molecular system by using, e.g.,  Jacobi coordinates the situation is much more involved for the rotational symmetry.
In this paper we address the latter problem using {\it holonomy reduction}. 
We suggest that the configuration space may be considered as the reduced holonomy
bundle with a connection induced by the mechanical connection.  Using the fact that for the special case of the  three-body problem, the holonomy group is $SO(2)$ (as opposed to  $SO(3)$ like in systems with more than three bodies)
we obtain a holonomy reduced configuration space of topology $ \mathbf{R}_+^3 \times S^1$.  The dynamics then takes place on the cotangent bundle over the holonomy reduced configuration space. 
On this phase space there is an $S^1$ symmetry action coming  from the conserved reduced angular momentum which can be reduced using the standard symplectic reduction method.
Using a  theorem by Arnold it follows that the resulting symmetry reduced phase space is again a natural mechanical phase space, i.e. a cotangent bundle. 
This is  different from what is obtained from the usual approach where  symplectic reduction is used from the outset. 
This difference is discussed in some detail, and a connection between the reduced dynamics of a triatomic molecule and the motion of a charged particle in a magnetic field is established. 
\end{abstract}

\noindent
AMS classification numbers: 70F07, 70G65, 53C80
\vspace{2pc}

\newpage

\section{Introduction}


In molecular dynamics which is the subject of this paper and generally in dynamical systems theory 
the reduction of the number of degrees of freedoms is of central importance
for both computational and  conceptual reasons.
A molecular system is a many body system consisting of the nuclei and electrons of the constituting atoms. 
The electronic degrees of freedoms are typically dealt with in a Born-Oppenheimer approximation.
Since the nuclear masses are a few thousands times bigger  than the mass of an electron one assumes that the nuclei adiabatically interact via the forces obtained from a potential 
energy surface that is obtained from the electronic ground state energy as a function of the nuclear configurations. 
The computation of such potential energy surfaces is based on density functional theory and other methods  and is an art in physical chemistry.
For several molecular systems such potential energy surfaces are tabulated in the chemistry literature.
Given such a potential energy surface a molecular system reduces to an $N$-body system which only involves the degrees of freedom of the $N$ nuclei in the system. 
This $N$-body system can then be treated classically or quantum mechanically. In particular for light atoms (respectively nuclei) like hydrogen quantum effects might play an important role which make a quantum mechanical treatment necessary. We note that state of the art quantum computations for, e.g., reactive scattering are even today only feasible for three or maximally four atoms. 
For this reason and also conceptual reasons one desires to get rid of as many degrees  as possible. 
A reduction of the number of `effective' degrees of freedom of a molecular $N$-body system can be achieved by exploiting the symmetries of the system. These symmetries consist of overall translations and rotations. 
The reduction of translational degrees of freedom is simple and can be achieved by using Jacobi coordinates or changing to a centre of mass coordinate system.
For rotations, the situation is much more involved as a clear distinction between rotational degrees of freedom and (internal) vibrational degrees of freedom only exists in an approximate sense in the vicinity of an equilibrium position.  Here the distinction between vibrations and rotations can be achieved from the so called  \textit{Eckart frame} \cite{Eckart35} that is widely used in 
applications \cite{WilsonDeciusCross55}.  This approximation is however only of local validity since large amplitude vibrations may produce rotations.
A major step towards a geometric understanding of why  a separation of rotations and vibrations cannot be achieved globally 
goes back to the work of  Guichardet \cite{Guichardet84} who used the differential geometry framework of
principal bundles to give a mathematically rigorous definition of vibrational motions. He showed that the
translational reduced configuration space is a principal bundle with
structure group given by the special orthogonal group, and introduced a connection which naturally relates to  molecular motions. 
The inseparability of rotations and vibrations then follows from the nonvanishing 
curvature of this so called \textit{mechanical connection}. 
Iwai and Tachibana \cite{IwaiTachibana86,IwaiTachibana99} used  Guichardet's approach to study in great detail both the classical and the quantum mechanical dynamics of $N$-body molecular systems.
Using the setting of principle bundles  Iwai\cite{Iwai87a} in particular showed that the Eckart frame can also be defined for general
configurations (i.e., no necessarily  equilibrium configurations) of a molecule. 
However, this frame is then not unique and therefore not suitable for studying large amplitude vibrational motions  
of a molecule. 
Iwai moreover applied the
Marsden-Weinstein-Meyer symplectic reduction procedure \cite{MarsdenWeinstein74, Meyer73} to 
reduce the constant angular momentum motion of an $N$-body system.
He showed that for nonvanishing angular momentum the reduced phase space is then no longer a natural mechanical system in the sense that it is no longer given as the cotangent bundle over a (reduced) configuration space. 
A gauge theoretical interpretation of the reduction of symmetries and the related choice of a reference frame in $N$-body systems was introduced 
in \cite{ShapereWilczek87,LittlejohnReinsch97}.
In their constructive and instructive paper Littlejohn and Reinsch\cite{LittlejohnReinsch97} used Lagrangian reduction instead of symplectic
reduction mentioned above. For more related work we mention the
refer to \cite{Iwai87b,IwaiYamaoka05,IwaiYamaoka08,YanaoKonMarsden06,Montgomery96,LittMitchAquiCava98}.


In this paper we use modern tools from the geometric description of molecular motion described  above
to introduce a new way to  reduce the symmetry specifically  of
triatomic molecular system. We obtain a reduced
configuration space and deduce the reduced dynamics for a triatomic molecule in a way which can be summarized as follows. Consider  three atoms (or nuclei) in  $\mathbf{R}^{3}$. 
Using  Jacobi coordinates the translational symmetry in the absence of external forces can be used to reduce the nine-dimensional configuration space
$\mathbf{R}^{3}\times \mathbf{R}^{3}\times 
\mathbf{R}^{3}$ of the triatomic system to the six-dimensional space  $\mathbf{R}^{3}\times \mathbf{R}^{3}$. 
Excluding collinear (and hence also collisional) configurations from
 $\mathbf{R}^{3}\times \mathbf{R}^{3}$ gives the translation reduced configuration space $P$ 
on which the special orthogonal
group $SO(3)$ acts freely. The space $P$ is a principal bundle with base
space given by the positive half space $\mathbf{R}_{+}^{3}$ \cite{Iwai87b}. Kinetic
energy gives a metric on $P$, and a connection can be obtained by defining
horizontal spaces as orthogonal complements of the tangent spaces of orbits
of the $SO(3)$ action. As known \cite{Guichardet84,LittlejohnReinsch97} the connection on $P$ has a nontrivial holonomy group
which is $SO(2)$.  This enables us  to use the holonomy reduction theorem \cite{KN} 
to reduce $P$ to the holonomy bundle which we denote by $Q$. Since $P$ is a trivial
bundle \cite{Iwai87b}, $Q$ is also trivial and hence topologically $\mathbf{R%
}_{+}^{3}\times SO(2)$, or equivalently $\mathbf{R}_{+}^{3}\times S^{1}$.  
The reduced phase space is then given by the cotangent bundle $T^{\ast} Q$. 
We explicitly derive the
Hamiltonian on $T^{\ast} Q$
and deduce the
reduced dynamics on $T^{\ast} Q$. 
In the final step we then use the conservation of the reduced angular momentum  
related to an $S^1$ action on $T^{\ast} Q$
to apply the
symplectic reduction procedure. Using a theorem in \cite{Arnold78} we find that
the reduced phase space is then a natural mechanical system, namely the cotangent bundle over 
 $Q/S^{1}$.

We note that there is no
natural way to generalize these results to
systems of four or more atoms. The reason is that 
triatomic systems are in many respects special. For example, the
holonomy group of a system of four or
more atoms is $SO(3)$, and the translation reduced space is not a
trivial bundle \cite{LittlejohnReinsch97}.

\section{Reduced configuration space}

\subsection{Principle bundle picture}

Consider a molecular system of three atoms. Let $\mathbf{x}_{i} \in \mathbf{R}$, $i=1,2,3$, be the
position vectors of these atoms. Suppose that there are no external forces. Then
the mass-weighted Jacobi vectors 
\begin{eqnarray*}
\mathbf{r}&=&\sqrt{\frac{m_{1}m_{2}}{m_{1}+m_{2}}}(\mathbf{x}_{1}-\mathbf{x}_{3}),\\
\mathbf{s}&=&\sqrt{\frac{m_{2}(m_{1}+m_{2})}{m_{1}+m_{2}+m_{3}}} 
(\mathbf{x}_{2}-\frac{m_{1}\mathbf{x}_{1}+m_{3}\mathbf{x}_{3}}{m_{1}+m_{3}}),
\end{eqnarray*}%
can be chosen to reduce the symmetry of overall translations. 
(For different choices of Jacobi vectors see \ref{sec:kinematicgroup}.) 
Excluding collinear (and hence also collisional) configurations 
we obtain the six-dimensional translation reduced configuration space 
\begin{equation*}
P=\left\{ x=(\mathbf{r},\mathbf{s}):\ \lambda \mathbf{r}+\mu \mathbf{s}\neq
0 \mbox{ for all } (\lambda ,\mu) \in \mathbf{R}^2\backslash \{0\}\right\} \subset \mathbf{R}^{3}%
\mathbf{\times R}^{3}.
\end{equation*}%
Proper rotations $g\in SO(3)$ act on $P$ in the natural way%
\begin{equation*}
g(\mathbf{r},\mathbf{s})=(g\mathbf{r},g\mathbf{s})\,.
\end{equation*}%
On $P$ this action is free and it thus follows from standard results that
\[
M:=P/SO(3) 
\]
has a manifold structure.
The space $M$  is usually referred to as \textit{shape space} or \textit{internal space}. 
Furthermore, the canonical projection $\pi :P\rightarrow M$ defines a principal bundle with structure group 
$SO(3)$ \cite{Guichardet84}. This means that $P$ consists of  smoothly glued copies
of $SO(3)$, i.e.,
locally, $P$ is diffeomorphic to $M\times SO(3)$.
Topologically, this local decomposition also holds  globally which following Iwai \cite{Iwai87b} can be seen as follows. 
Using Jacobi coordinates 
\begin{equation*}
r=\sqrt{\left\langle \mathbf{r},\mathbf{r}\right\rangle },\ s=\sqrt{%
\left\langle \mathbf{s},\mathbf{s}\right\rangle },\ \phi =\cos ^{-1}\left(
\left\langle \mathbf{r},\mathbf{s}\right\rangle /rs\right) ,
\end{equation*}%
where $\left\langle \cdot ,\cdot \right\rangle $ is the usual dot product on $\mathbf{R}^{3}$,
and introducing coordinates%
\begin{equation*}
w_{1}=r^{2}-s^{2},\ w_{2}=2rs\cos \phi ,\ w_{3}=2rs\sin \phi >0
\end{equation*}%
one sees  that 
$M\cong \mathbf{R}_{+}^{3}=\left\{ (w_{1},w_{2},w_{3}): w_{3}>0\right\} $. As pointed out in \cite{Iwai87b}, $P$ is a
trivial bundle as $M$ is contractible to a single point. So, topologically, 
$P\cong \mathbf{R}_{+}^{3}\times SO(3)$.

\subsection{Nontrivial holonomy} 

Turning back to the action of $SO(3)$ on $P$ one can see that the 
\textit{fundamental vector field} $\widetilde{A}$ associated with an element $A$ in the Lie algebra $so(3)$ is given by%
\begin{equation}
 \widetilde{A} |_x  =  \left.\frac{d}{dt} \right |_{t=0}(e^{tA}x)\,,
\end{equation}%
or  equivalently, %
\begin{equation}
\widetilde{A}  |_x  =(A\mathbf{r},A\mathbf{s})=(\mathbf{w}\times \mathbf{r},%
\mathbf{w}\times \mathbf{s}),  \label{fun}
\end{equation}%
where $\mathbf{w}\in \mathbf{R}^{3}$ is the unique vector corresponding to 
$A $ by the  isomorphism 
\begin{equation}
R^{-1}:so(3)\rightarrow \mathbf{R}^{3}, \quad 
\left(
\begin{array}{ccc}
0 & -a_3 & a_2 \\ 
a_3 & 0 & -a_1 \\ 
-a_2 & a_1 & 0
\end{array}
\right) \mapsto
\left(
\begin{array}{ccc}
a_1 \\ 
a_2  \\ 
a_3 
\end{array}
\right)\,. \label{eq:def_R}
\end{equation} 
Let $N$ be an orbit of the $SO(3)$ action, say $N=SO(3)x$ for a point $x\in
P $, then $T_{x}N=\left\{ \widetilde{A} |_x : A\in so(3) \right\} $. 
Consider the orthogonal complement $H_{x}$ of $T_{x}N$ in $T_{x}P $ with respect to the Euclidean dot product on $P$ given by%
\begin{equation}
dx^{2}=\left\langle \mathbf{r},\mathbf{r}\right\rangle +\left\langle 
\mathbf{ s},\mathbf{s}\right\rangle .  \label{metric}
\end{equation}%
Clearly the distribution $x\mapsto H_{x}$, which we call the \textit{horizontal distribution},
defines a connection \cite{Iwai87b} $\omega :TP\rightarrow so(3)$
on $P$ which is a special case of the \textit{mechanical connection} defined in
\cite{Marsden92}. A vector field $X^{\ast }$ with $X^{\ast } |_x \in H_{x}$
for all $x\in P$ is called \textit{horizontal}.  The 
\textit{horizontal lift} of a vector field $X$ on $M$ is accordingly the unique horizontal vector field 
$X^{\ast }$ on $P$ such that $d\pi (X^{\ast })=X$. We have $\omega
(X^{\ast })=0$ for every horizontal vector field $X^{\ast }$ and $\omega (%
\widetilde{A})=A$ for every fundamental vector field $\widetilde{A}$. In
order to compute the horizontal lifts of the coordinate vector fields $%
\partial _{r},\partial _{s},\partial _{\phi }$ on $M$ we give an explicit expression for the metric  $dx^{2}$
in (\ref{metric}). To this end we follow  \cite{LittlejohnReinsch97,IwaiYamaoka08}
and introduce a frame $\mathbf{u}_{1},
\mathbf{u}_{2},\mathbf{u}_{3}$ in $\mathbf{R}^{3}$ according to
\begin{eqnarray*}
\mathbf{r} &=&r \, \mathbf{u}_{1}, \\
\mathbf{s} &=&s\, \cos \phi \mathbf{u}_{1}+s\sin \phi \,  \mathbf{u}_{2}, \\
\mathbf{u}_{3} &=&\mathbf{u}_{1}\times \mathbf{u}_{2}.
\end{eqnarray*}%
If Euler angles ($\alpha ,\beta ,\gamma $) on $SO(3)$ are chosen via
\begin{equation*}
g=e^{R(\alpha \mathbf{e}_{1})}e^{R(\beta \mathbf{e}_{2})}e^{R(\gamma \mathbf{%
e}_{3})},\ 0\leq \alpha ,\gamma \leq 2\pi ,\ 0\leq \beta \leq \pi ,
\end{equation*}%
where $\mathbf{e}_{1},\mathbf{e}_{2},\mathbf{e}_{3}$ is the standard basis
of $\mathbf{R}^{3}$, $R$ is defined in (\ref{eq:def_R})  and $g\mathbf{e}_{i}=\mathbf{u}_{i}$, $i=1,2,3$, then with
\begin{eqnarray*}
\Theta _{1}&=&\sin \gamma \, d\beta -\sin \beta \cos \gamma \, d\alpha ,\\
\Theta _{2}&=&\cos \gamma \, d\beta +\sin \beta \sin \gamma \, d\alpha ,\\
\Theta _{3}&=&\cos \beta \, d\alpha +d\gamma, 
\end{eqnarray*}%
one obtains \cite{IwaiYamaoka08}%
\begin{eqnarray*}
d\mathbf{r} &=&dr\,\mathbf{u}_{1}+r\Theta _{3}\mathbf{u}_{2}-r\Theta _{2}\mathbf{u%
}_{3}, \\
d\mathbf{s}&=&\eta _{1}\mathbf{u}_{1}+\eta _{2}\mathbf{u}_{2}+\eta _{3}\mathbf{%
u}_{3},
\end{eqnarray*}%
where
\begin{eqnarray*}
\eta _{1}&=&ds\cos \phi -s\sin \phi \, d\phi -s\sin \phi \, \Theta _{3}, \\
\eta _{2}&=&ds\sin \phi +s\cos \phi \, d\phi +s\cos \phi \, \Theta _{3},\\
\eta _{3}&=&s\sin \phi \, \Theta _{1}-s\cos \phi \, \Theta _{2}.
\end{eqnarray*}
In local coordinates the metric $d\mathbf{x}^{2}$ then assumes the form
\begin{equation*}
d\mathbf{x}^{2}=dr^{2}+r^{2}(\Theta _{2}^{2}+\Theta _{3}^{2})+\eta
_{1}^{2}+\eta _{2}^{2}+\eta _{3}^{2}.
\end{equation*}%
This expression can be used to locally compute the horizontal lift $X^{\ast }$ of a
vector field $X$ on $M$: $X^{\ast }$ is orthogonal to $\partial _{\alpha
},\partial _{\beta },\partial _{\gamma }$, and $d\pi (X^{\ast })=X$. It
follows that
\begin{equation}
\partial _{r}^{\ast }=\partial _{r},\ \partial _{s}^{\ast }=\partial _{s},\
\partial _{\phi }^{\ast }=\partial _{\phi }-\frac{r^{2}}{r^{2}+s^{2}}%
\partial _{\gamma }.  \label{hor}
\end{equation}%
In gauge theory the factor $\frac{r^{2}}{r^{2}+s^{2}}$ is referred to as a component of a \textit{Yang-Mills potential} 
\cite{IwaiTachibana99}.

\subsection{Holonomy reduction}

By Equation~(\ref{hor}) we have arrived at the well-known phenomena of inseparability of rotations
and vibrations \cite{Guichardet84,Iwai87a}. Namely from (\ref{hor}) we see 
that the distribution spanned by 
$\partial _{r}^{\ast },\partial _{s}^{\ast},\partial _{\phi }^{\ast }$ is not integrable, and hence, if these vector fields are
considered as infinitesimal vibrational motions one can say that
vibrations generate rotations. This is why the internal space $M$ is not a
submanifold of $P$ \cite{IwaiTachibana99}. On the other hand $\partial
_{r}^{\ast },\partial _{s}^{\ast },\partial _{\phi }^{\ast },\partial
_{\gamma }$ \emph{do} span an involutive and hence integrable distribution. The
maximal integral manifold $Q_{x}$ of that distribution at a point $x\in P$
is a good candidate for being the reduced configuration space because
vibrational motions through $x$ live in that space. 
In fact we will obtain the reduced dynamics of a triatomic molecule on the cotangent bundle over $Q_x$
by employing the   holonomy reduction of principle
bundles:  
A curve on $P$ is called horizontal if its tangents are horizontal.
Fix a point $x\in P$ and denote by $P(x)$ the set of all points in $P$ which can
be joined to $x$ by horizontal curves. It is known that \cite{Guichardet84,LittlejohnReinsch97} the holonomy group of $\omega $ is $SO(2)$ 
(see also  \ref{sec:Guichardet}), and since $M$ is connected and paracompact the holonomy reduction
theorem \cite{KN} implies that $P(x)$ is a reduced bundle with structure group $SO(2)$, 
which is in fact $Q_{x}$. Furthermore, $Q_{x}$ is a trivial bundle as it
has the same base space as $P$. These observations suggest that the reduced
configuration space of a triatomic molecular system is topologically 
$\mathbf{R}_{+}^{3}\times SO(2)$.  The induced metric on $Q_{x}$ is thus%
\begin{equation}
dq^{2}=dr^{2}+ds^{2}+\frac{r^{2}s^{2}}{r^{2}+s^{2}}d\phi ^{2}+\frac{1}{%
r^{2}+s^{2}}\zeta ^{2},  \label{redmet}
\end{equation}%
where%
\begin{equation}
\zeta =s^{2}d\phi +(r^{2}+s^{2})d\gamma. \label{form}
\end{equation}

\section{Reduced dynamics}

\subsection{Angular momentum}

In the following we want to put our derivation above into the context of some well known results.
It is known  \cite{Eckart35}  that in  the case of small vibrations one can
separate vibrations and rotations in the vicinity of an equilibrium point.
In the present situation if one chooses $d\gamma =0$ in (\ref{redmet}) the
well-known Eckart kinetic energy is obtained. This is the gauge dependent
internal metric $h_{\mu \nu }$ in \cite{LittlejohnReinsch97}. Thus one can
conclude that in case of small vibrations the internal motions of molecule
live in the integral manifolds of the distribution spanned by $\partial
_{r},\partial _{s},\partial _{\phi }$, called the \textit{Eckart space}.
Next, consider the angular momentum%
\begin{equation*}
\mathbf{J}=R(\mathbf{r}\times d\mathbf{r}+\mathbf{s}\times d\mathbf{s})
\end{equation*}%
on $P$ which is computed locally to be 
\begin{equation*}
\mathbf{J}=R((r^{2}\Theta _{2}-s\cos \phi \eta _{3})\mathbf{u}%
_{2}+(r^{2}\Theta _{3}+s\cos \phi \eta _{2}-s\sin \phi \eta _{1}+s\sin \phi
\eta _{3})\mathbf{u}_{3}).
\end{equation*}%
So, its restriction to $Q_{x}$ is%
\begin{equation*}
\mathbf{J} |_{_{Q_{x}}} =R(\zeta \mathbf{u}_{3}).
\end{equation*}%
If the angular momentum of the system is identically zero, then $\zeta
=0$, and hence%
\begin{equation*}
dq^{2}=dr^{2}+ds^{2}+\frac{r^{2}s^{2}}{r^{2}+s^{2}}d\phi ^{2}.
\end{equation*}%

\subsection{Holonomy reduced Hamiltonian}

In the case of vanishing angular momentum  the Hamiltonian is obtained to be%
\begin{equation}
H=\frac12  p_{r}^{2}+ \frac12 p_{s}^{2}+ \frac12 (\frac{1}{r^{2}}+\frac{1}{s^{2}})p_{\phi
}^{2}+V(r,s,\phi ),  \label{ham}
\end{equation}
where $p_{r},p_{s},p_{\phi }$ are the conjugate momenta and $V(r,s,\phi )$
is the potential energy which is assumed to be rotationally invariant. This
Hamiltonian is widely used in applications. By (\ref{form}) we observe that
vibrational motions live in the integral manifold of the distribution
spanned by $\partial _{r}^{\ast },\partial _{s}^{\ast },\partial _{\gamma }$.
That space may be called \textit{zero-angular momentum space}.

If the reduced angular momentum $\mathbf{J} |{_{Q_{x}}}$ is a non-zero constant,
we have $\zeta=const.\ne 0$. Then, equivalently, $\partial _{\phi }^{\ast }$ is a non-zero constant
and hence the vibrational motions remain in a three-dimensional affine space which is parallel
to the zero-angular momentum space. 

Taking into account the contribution of $\zeta$ in the induced metric $dq$ on $Q_x$ in (\ref{redmet})  the Hamiltonian in (\ref{ham}) changes to the general
\textit{holonomy reduced Hamiltonian}  
\begin{equation*}
H=\frac{1}{2}p_{r}^{2}+\frac{1}{2}p_{s}^{2}+\frac{1}{2}(\frac{1}{r^{2}}+%
\frac{1}{s^{2}})p_{\phi }^{2}-\frac{1}{r^{2}}p_{\phi }p_{\gamma }+\frac{1}{%
2r^{2}}p_{\gamma }^{2}+V(r,s,\phi ).
\end{equation*}
The corresponding Hamiltonian vector field is given by
\begin{eqnarray*}
X=& &p_{r}\partial_r+ p_{s} \partial_s +
\left( (\frac{1}{r^{2}}+\frac{1}{s^{2}})p_{\phi }  -\frac{1}{r^2} p_\gamma \right) \partial_\phi +
\frac{1}{r^{2}} (p_{\gamma }-p_{\phi }) \partial_\gamma + \\
& & \left( \frac{1}{r^{3}}(p_{\gamma }-p_{\phi })^{2} - \frac{\partial V}{\partial r} \right) \partial_{p_r} +
\left(\frac{1}{s^{3}}p_{\phi }^{2}-\frac{\partial V}{\partial s}  \right) \partial_{p_s} - 
\frac{\partial V}{\partial \phi }  \partial_{p_\phi} \,.
\end{eqnarray*}

Since $\gamma$ is cyclic the conjugate momentum $p_\gamma$ is conserved. To put it another way $\mathbf{J} |_{_{Q_{x}}} $ is an $S^1$-equivariant momentum 
and the standard symplectic reduction theorem can be applied. Using 
a theorem by Arnold (see  \cite{Arnold78}, page 378) the resulting reduced phase space is again  a natural mechanical system, i.e. a cotangent bundle.

\section{Comments on related work}

\subsection{The relation between the motions of a triatomic molecule and a charged particle in a magnetic field}

In \cite{Montgomery90, MarsdenRatiu99} the idea is introduced to describe the motion of a charged particle in a magnetic field by extending the configuration space 
$\mathbf{ R}^3$ to $\mathbf{ R}^3 \times S^1$ such that the angle corresponding to $S^1$ is cyclic and its conserved conjugate momentum gives the charge of the particle in the magnetic field.
Since the  holonomy reduced configuration space is $\mathbf{R}_{+}^{3}\times S^{1}$ we can identify the motion of a triatomic molecule to that of a charged particle in a magnetic field as follows.  
Let  $\mathbf{q}$ denote a point in $M$ with coordinates $(r,s,\phi )$. If   $\mathbf{A}$ denote the one-form
$(r^{2}+s^{2})^{-1}\zeta =
\frac{s^{2}}{(r^{2}+s^{2})}d\phi +d\gamma $ on $\mathbf{ R}_{+}^{3}$, then by the metric (\ref{redmet}) the kinetic energy can be written as
\[
L_{K}=\frac{1}{2}\left\Vert \mathbf{\dot{q}}\right\Vert ^{2}+\frac{1}{2}(%
\mathbf{A}\cdot \mathbf{\dot{q}}+\dot{\gamma})^{2}
\]%
which is reminiscent of the so called \textit{Kaluza-Klein Lagrangian} 
\cite{MarsdenRatiu99}. The conjugate momenta are then
\[
\mathbf{p}  =  \frac{\partial  L_{K}}{\partial \mathbf{\dot{q}}} = \mathbf{\dot{q}}+(\mathbf{A}\cdot \mathbf{\dot{q}}+\dot{\gamma})%
\mathbf{A}
\]%
and%
\[
p_\gamma =   \frac{\partial  L_{K}}{\partial \dot{\gamma}} = \mathbf{A}\cdot \mathbf{\dot{q}}+\dot{\gamma}.
\]%
The one-form   $\mathbf{A}$ plays the role of a vector potential for the magnetic field. The conserved momentum 
$p_\gamma$ is the charge $e=c p_\gamma$ (with $c$ denoting the speed of light) \cite{MarsdenRatiu99}.

\subsection{Relation to symplectic and dimensional reduction}

In \cite{Iwai87a} the symplectic reduction procedure \cite{MarsdenWeinstein74, Meyer73} is applied to the $N-$body problem. The
cotangent bundle over the translation reduced configuration space $P$ is a symplectic
manifold with the canonical two-form, and the angular momentum $J:T^{\ast
}P\rightarrow so(3)$ is an equivariant momentum map. For a $\mu \neq 0$, it
is shown that $J^{-1}(\mu )$ is a principal bundle with structure group 
$SO(2)$ whereas the zero momentum space $J^{-1}(0)$ is a principal bundle with
structure group $SO(3)$. Furthermore $J^{-1}(0)/SO(3)$ is shown to be
diffeomorphic to $T^{\ast }(P/SO(3))$, but $J^{-1}(\mu )/SO(2)$ is no more a
cotangent bundle because of dimensionality. As pointed out in \cite{Iwai87a}
the procedure for the latter when applied to for three-bodies is in fact the 
\textit{elimination of nodes}.

In contrast to the symplectic reduction procedure the first step
in this paper was to pass from the translation reduced configuration space $P$ of a triatomic molecule to a subbundle $Q$ 
(the holonomy reduced bundle) which is a principle bundle with structure group $SO(2)$. 
Afterwards the angular momentum is then restricted to $T^{\ast }Q$, and finally
the sypmlectic reduction procedure is applied.  The  reduced  space is then always a
cotangent bundle as follows from a theorem by Arnold (see  \cite{Arnold78}, page 378). 

We note that  the method used in the present work is strongly related to
dimensional reduction \cite{ForgacsManton80, ShniderSternberg83}, a method
developed for symmetries of gauge fields. More precisely, in the case of spherical symmetry in 6 dimensions applied to an SU(3) gauge theory, 
the 2 extra dimensions describing a sphere of radius $R$. One solution, with the largest set of Higgs fields, reduces to the 4-dimensional 
Weinberg-Salam model without fermions \cite{ForgacsManton80}.

\section{Conclusions}

In this paper we used the geometric theory of molecular mechanics  \cite{Guichardet84, Iwai87b, LittlejohnReinsch97}  
to reduce the number of degrees of freedom in the molecular three-body problem. We followed the principal bundle setting of Guichardet\cite{Guichardet84} on the translation reduced configuration space, and using the holonomy reduction theorem \cite{KN} it was possible to reduce to a principle subbundle. This may be interpreted as separating two rotational degrees of freedom from the maximal space that includes vibrational motions. It was then possible to induce the angular momentum and apply the very symplectic reduction procedure (to be precise, we used it in the form of the Noether's theorem here). 
This way, the remaining momentum space which is of 6 dimensions and also a phase space was obtained. The computations were local for the purpose illustration but the method is intrinsic. 
In some sense, the resulting space which is of 3 dimensions can be seen as a rotationless space. 
For the case of zero angular momentum this is a known earlier result. In the present paper it was generalized to the case of non-zero angular momentum. 
In particular we used our approach to rephrase the well known fact \cite{Montgomery90} that a triatomic molecular system behaves as a single particle in a magnetic field. 


\appendix

\section{A lemma by Guichardet}
\label{sec:Guichardet}

For completeness, we give a brief proof of the fact in that a vibrational
motion of a triatomic molecule, which is defined as a curve with horizontal tangents, remains in a fixed plane as originally formulated  by Guichardet
\cite{Guichardet84}: Let $x(t)=(\mathbf{r}(t),\mathbf{s}(t))$ be a horizontal curve
on $P$. We  show that 
$F_{x(t)}:=\mbox{span}\left\{ \mathbf{r}(t),\mathbf{s}(t)\right\} $ is fixed. 
Indeed, since $x(t)$ is horizontal it is orthogonal to all
fundamental vector fields which are given in (\ref{fun}), and hence $\mathbf{r}%
(t)\times \mathbf{\dot{r}}(t)+\mathbf{s}(t)\times \mathbf{\dot{s}}(t)=0$.
Let $\mathbf{y}(t)$ be a curve in $\mathbf{R}^{3}$ with $\left\langle 
\mathbf{y}(t),\mathbf{y}(t)\right\rangle =1$ which is orthogonal to $%
F_{x(t)} $. So, $\mathbf{\dot{y}}(t)$ is orthogonal to $F_{x(t)}$. Hence $%
\left\langle \mathbf{y}(t),\mathbf{r}(t)\right\rangle =\left\langle \mathbf{y%
}(t),\mathbf{s}(t)\right\rangle =0$ so $\left\langle \mathbf{\dot{y}}(t),%
\mathbf{r}(t)\right\rangle =\left\langle \mathbf{\dot{y}}(t),\mathbf{s}%
(t)\right\rangle =0$ which implies $\mathbf{\dot{y}}$ is zero.

As a conclusion of the above fact it is observed  
\cite{LittlejohnReinsch97} that during vibrational motions or shape deformations the Jacobi
vectors remain in a fixed plane, and hence the  Jacobi vectors before and after the vibrational motion can be
transformed to one another by a plane rotation, i.e. the holonomy
group is $SO(2)$.

\section{The kinematic group}
\label{sec:kinematicgroup}

Different clusterings of position vectors give rise to different choices of
mass-weighted Jacobi vectors. These different choices are related to each other
 by transformations which are called \textit{democracy transformations}
 \cite{LittlejohnReinsch97}. The set of all democracy transformations forms a
subgroup of the symmetry group $SO(3)$ called \textit{democracy} or 
\textit{kinematic group}. For the three-body problem the kinematic group is $SO(2)$. This is another special feature of the three-body problem.


\end{document}